\documentclass[
aps,nofootinbib,
showpacs,showkeys,preprint
tightenlines,preprintnumbers,] {revtex4}

\usepackage{epsf,epsfig,subfigure,graphicx,amsmath,amssymb 
}
\usepackage{color}
\newcommand{\dis}[1]{\begin{equation}\begin{split}#1\end{split}\end{equation}}
\newcommand{\ie}{{\it i.e.~}}

\newcommand{\tev}{\,\textrm{TeV}}
\newcommand{\gev}{\,\textrm{GeV}}
\newcommand{\MeV}{\,\textrm{MeV}}

\newcommand{\eV}{\,\textrm{eV}}
\newcommand{\Mp}{M_{\rm P}}

\newcommand{\vew}{v_{\rm ew}}

\newcommand{\Uanom}{U(1)$_{\rm anom}$}

\def\sw0{{$\sin^2\theta_W^0$}}

\def\Nf2{{\bf N_{[2]}}}

\newcommand{\Z}{{\bf Z}}

\def\E6{{\rm E_6}}

\def\EE8{{\rm E_8\times E_8'}}

\def\SU21{SU$(2)\times$U(1)}

\def\one{{\bf 1}}

\def\two{{\bf 2}}
\def\five{{\bf 5}}
\def\ten{{\bf 10}}
\def\tenb{{\overline{\bf 10}}}

\def\fiveb{{\overline{\bf 5}}}

\begin{document}

\draft

\title{\Large\bf  A model of dynamical SUSY breaking}

\author{Jihn E.  Kim$^{(1)}$, Bumseok Kyae$^{(2)}$}
\affiliation
{$^{(1)}$Department of Physics, Kyung Hee University, 26 Gyungheedaero, Dongdaemun-Gu, Seoul 02447, Republic of Korea,\\ 
$^{(2)}$Department of Physics, Pusan National University, 2 Busandaehakro-63-Gil, Geumjeong-Gu, Busan 46241, Republic of Korea 
}
 
\begin{abstract} 
 Supersymmetric (SUSY) models and  dynamical   breaking of symmetries have been used to explain hierarchies of mass scales.   We find that a chiral representation,  $\tenb\, \oplus\, \fiveb\, \oplus\,  2\cdot \five$   in SUSY SU(5) in the hidden sector, breaks  global SUSY dynamically, by producing a  composite field $\phi$ below the SU(5) confinement scale. This dynamincal SUSY breaking can have two  important applications, one  in particle physics and the other in cosmology.
Gavitational effects transmit this dynamical breaking  to the standard model(SM) superpartners and the quintessential vacuum energy. The SM superpartners feel the effects just by the magnitude of the gravitino mass while the smallness of the quintessential vacuum energy is due to the composite nature of a singlet field $\phi$.  The composite $\phi$ carries a global charge which is hardly broken in SUSY and hence its phase can be used toward a quintessential axion for dark energy of the Universe. 
 
\end{abstract}
\maketitle

\section{Introduction}\label{sec:Introduction}

Mass scales are the most important physical parameters in all physics disciplines.  In any theory, the definition of mass scale  is given by the Planck mass $\Mp$. The next important mass scale  determines a physics discipline.  It is a mystery how this next scale  has arisen from the fundamental Planck mass. In the standard model (SM) of particle physics, the next scale is the electroweak scale $\vew=246\,\gev$. In the standard Big Bang cosmology, the next scale is the current value of the Hubble parameter $H_0$. In  the current dark energy (DE) dominated Universe, DE can  be considered as the next scale, which we follow here.  In nuclear physics, atomic physics, and condensed matter physics, the next scales are $7\,\MeV, 1\,\eV, 10^{-3\,}\eV$, respectively, which can be derived in principle from the electroweak scale.  
Cosmological dark matter has its root in particle species \cite{KimBook18} and DE looks for solutions in the framework of general relativity \cite{Tsujikawa18}. Therefore, understanding mass parameters in two standard models, in particle physics and in cosmology, is the key in understanding all physical mass parameters.

Gauge forces in the SM can be unified into a grand unified theory (GUT) at the scale $10^{15-17\,}\gev$ \cite{GG74}. Understanding the ratio of the electroweak scale and the GUT scale is known as the gauge hierarchy problem \cite{Weinberg75}. If this hierarchy of masses is not fine-tuned, the best ideas so far suggested are behind two pillars: dynamical symmetry breaking(DSB)  \cite{Susskind79} and breaking of $N=1$ supersymmetry(SUSY)  \cite{Witten81}. In particular, the DSB idea is most natural in the sense that an exponentially small mass parameter can be obtained by the evolution of gauge couplings of the confining force. If the confinement is working for generating small mass scales, then it can generate even very small mass parameter such as the DE scale of $10^{-3\,}\eV$ because a composite particle with a large enegineering dimension may be the source of this DE scale.

The $N=1$ SUSY idea accompanies superpartners of the SM particles around the TeV scale. Once the small electroweak scale is introduced as a parameter at the GUT scale then it can be used down at the electroweak scale via the non-renormalization theorem  but it lacks in explaining the exponentially small scale itself.  Currently,  TeV scale SUSY particle effects have not shown up at the Large Hadron Collider, and we need to raise the superparticle masses above a few TeV   \cite{ATLASSUSY18,CMSSUSY19}.  By  raising the superpartner masses  above the TeV scale, there must be a kind of small fine-tuning of parameters, the so-called the little hierarchy problem (a problem on the SUSY particle masses and the electroweak scale), which can be understood now in various methods \cite{Kyae18}.

Since SUSY must be broken, generating a SUSY breaking scale is the key in SUSY solutions of gauge hierarchy problem. In this regard, a DSB in SUSY models  is  promising toward understanding hierarchical  mass scales. Phenomenologically, the SUSY idea must relate the electroweak scale to the SUSY breaking scale, which was popular in the early 1980's under the name of gravity mediation \cite{Nilles82,Barbieri82}. Supersymmetry breaking by gluino condensation needs the gluino condensation scale of order $10^{13\,}\gev$ \cite{GauginoFGN83} to obtain TeV scale gravitino mass.
Since it is required to raise the gravitino mass above several TeV  \cite{ATLASSUSY18,CMSSUSY19}, the gluino condensation idea may not work if the hidden sector and the visible sector gauge couplings are unified at the GUT scale. On the other hand, if an F-term of chiral field is the source for SUSY breaking, then the intermediate mass scale about $5\times 10^{10\,}\gev$ is needed for  gravitino mass of order the TeV scale \cite{Zumino77}.    In addition, this intermediate scale has been useful for the ``invisible'' axion and the $\mu$-term in supergravity  \cite{KimPLB84,KimNilles84}. 

We find that  a chiral representation,  $\tenb\, \oplus\, \fiveb\, \oplus\,  2\cdot \five$   in SUSY SU(5) in the hidden sector, breaks  global SUSY dynamically, by producing a  composite field $\phi$ below the SU(5) confinement scale $\Lambda$. The composite field $\phi$ respects a global symmetry U(1), which is shown by matching anomalies above and below the confinement scale. The F-term scale of $\phi$  can be   larger than $5\times 10^{10\,}\gev$, raising the masses of SM superpartners much above TeV. Since $\phi$ is composite, breaking the resulting global U(1) of $\phi$ can be very tiny, which can be the source of DE scale in the Universe \cite{KimBook18}.

\color{black}

\section{Dynamical breaking in  a SUSY GUT}

Early ideas on supersymmetric QCD(SQCD) in SU($N_c$) gauge group \cite{Davis83, Affleck84, Seiberg94PRD} are DSB models, which    are summarized in \cite{Amati88,Seiberg94PRD}. Our model here does not belong to this kind of SQCD but to a kind of a chiral-family model in SUSY grand unification (SGUT). The first study on DSB in SGUT was presented in \cite{Meurice84}, where  one chiral family in the SU(5) SGUT was suggested to induce SUSY breaking based on the flat direction argument. In the one-family SGUT SU(5), however, a superpotential cannot be written and only an argument toward SUSY breaking has been presented. 

In this paper, we find a chiral representation,  $\tenb\, \oplus\, \fiveb\, \oplus\,  2\cdot \five$, in  SU(5) SGUT with superpotentials  given above and below the confinement scale and attribute the source of SUSY breaking in supersymmetric SMs (SSMs)  to the confinement scale $\Lambda$ of this  hidden sector SU(5).   
For a non-Abelian gauge symmetry SU(5) and a non-Abelian global symmetry SU(2),  the representations are written as, (SU(5)$_{\rm gauge}$, SU(2)$_{\rm global}$), and we introduce
\dis{
& \bar{\Psi}=(\tenb,\one),~\bar{\psi}_1=(\fiveb,\one),  {\psi}_2=(\five,\two),~  D_1\sim(\one,\two) .\label{eq:qrepresentations}
}
We are guided to study Eq. (\ref{eq:qrepresentations}) from our earlier work \cite{Huh09}. These fields, being complex, can have phases. So, the full global symmetry we consider is SU(2)$\times $U(1)$_{\bar{\Psi} }\times$U(1)$_{\bar{\psi}_1 }\times$U(1)$_{{\psi}_2 }\times$U(1)$_{{D}_1 }$. Note that our global symmetry is complete.

There are three SU(2) invariant terms in the superpotential,\footnote{Two family SU(5) Georgi-Glashow model \cite{GG74},  $2( \ten\oplus\fiveb)$ in Ref. \cite{Amati88}, is different from Eq. (\ref{eq:qrepresentations}).}
\dis{
W_0 \ni \frac14 \bar{\Psi}^{\alpha\beta}\psi_{2\alpha}^i \psi_{2\beta}^j\epsilon_{ij},~\bar{\psi}_1^\alpha \psi_{2\alpha}^i D_{1i},~\frac{1}{5!}\bar{\Psi}^{\alpha\beta}\bar{\Psi}^{\gamma\delta}\bar{\psi}_1^{\epsilon}\,\epsilon_{\alpha\beta\gamma\delta\epsilon}, \label{eq:Wterms} 
}
which act as three constraints on the U(1) phases,\footnote{Here, we express charges as U(1)'s.} 
\dis{
&U(1)_{\bar{\Psi} } + 2U(1)_{\psi_2}=0,\\
&U(1)_{\bar{\psi}_1 } +U(1)_{\psi_2}+U(1)_{D_1}=0,\\
&2U(1)_{\bar{\Psi} } +U(1)_{\bar{\psi}_1 }=0.
}
Therefore, there remains only one U(1) global symmetry, say $U(1)_{\bar{\Psi} }$, and the other phases can be written as, 
$
U(1)_{\bar{\psi}_1 }=-2U(1)_{\bar{\Psi} },~ U(1)_{\psi_2}=-\frac12 U(1)_{\bar{\Psi} },$ and $ U(1)_{D_1}=\frac52 U(1)_{\bar{\Psi} }.
$
Global symmetries survive below the confinement scale, where only SU(5) singlets will be considered.  
If the conefinement preserves SUSY and the confining SU(5)$_{\rm gauge}$ is not broken,  we can consider the following SU(5)$_{\rm gauge}-$singlet chial fields,
\dis{
\phi_1=\frac{1}{5!}\bar{\Psi}^{\alpha\beta}\bar{\Psi}^{\gamma\delta}\bar{\psi}_1^{\epsilon}\,\epsilon_{\alpha\beta\gamma\delta\epsilon},~  \phi_2=\frac14\bar{\Psi}^{\alpha\beta}\psi_{2\alpha}^i \psi_{2\beta}^j\epsilon_{ij},~\Phi_i=\bar{\psi}_1^\alpha {\psi_{2\,\alpha\,i}} .\label{eq:Comp}
}

\begin{table}[t!]
\begin{center}
\begin{tabular}{@{}|c|c|c|cccc|cc|c||c|@{}} \toprule
 &$2\ell(R_{SU(5)})$&  $SU(2)$ &$U(1)_{\bar{\Psi} }$& $U(1)_{\bar{\psi}_1 }$&$U(1)_{\psi_2}$ &$U(1)_{D_1}$  &$U(1)_{AF}$ &$U(1)_{R}$& dimension  
  \\[0.2em] \colrule &&&&&&&&& \\[-1.2em] 
 $\vartheta$  & $0$  & $0$ &   $0$ & $0$& $0$& $0$&  $0$ &  $+1$&$\frac12$ \\
  [0.2em]
$\bar{\Psi} \sim(\tenb,\one)$  & $-$  & $\one$ &$+1$ & $0$ & $0$ &   $0$ &  $-1$&  $+1 $ &  $1$ \\
  [0.2em]
$\qquad  {\rm fermion}$  & ${ +3}$  & $\one$ &$+1$ & $0$&$0$ & $0$ &   $-1$  &  $0$ &$-$ \\
  [0.2em]
 $\bar{\psi}_1\sim(\fiveb,\one)$&$-$& $\one$  & $0$  &  $+1$ & $0$ & $0$ &  $+2$ &   $0$ &$1$  \\
  [0.2em]
$\qquad  {\rm fermion}$  & $  +1$ &  $\one$ &   $0$ & $+1$ &$0$& $0$ &$+2$   &  $-1$&$- $   \\
  [0.2em]
$ {\psi}_2\sim(\five,\two)$  & $-$   & $\two$  & $0$ & $0$ &   $+1$   & $0$ &  $+\frac12$&  $+\frac{1}{2}$ & $1$\\
  [0.2em]
$\qquad  {\rm fermion}$  & $+1\times 2$   & $\two$ & $0$  & $0$ &   $+1$  & $0$  &  $+\frac12$&  $-\frac{1}{2}$ &$-$\\[0.6em]
 $ D\sim(\one,\two)$  & $-$   & $\two$  & $0$ & $0$   & $0$ &   $+1$ &  $-\frac52$ &  $+\frac{3}{2}$ &$1$\\
  [0.2em]
$\qquad  {\rm fermion}$  & $+1\times 2$   & $\two$ & $0$  & $0$& $0$    & $+1$  &  $-\frac52$ &  $+\frac{1}{2}$ &$-$\\[0.6em]
 $W^a\sim \lambda^a$&$0$  & $-$  & $0$ & $0$ & $0$ & $0$ & $0$ &    $+1$ &$\frac32$ \\
  [0.2em]
 $\Lambda^b$ & & $-$& $-$   & $-$ & $-$ & $-$ &  $-$ &   $\frac{2b}{3}$& $b$  \\[0.2em] 
  \hline 
 $\phi$    & $-$  & $\one$   & $-$ & $-$ & $-$ &$-$&  $-5$&$+2$ &$1$  \\
  [0.2em]
$\qquad  {\rm fermion}$  & $-$   & $\one$  & $-$ & $-$  & $-$  & $-$&  $-5$& $+1$ &$-$ \\ [0.2em]
 $\Phi_i$ & $-$   & $\two$ & $-$   & $-$ & $-$ &$-$ &  $+\frac52$ & $+\frac{1}{2}$& $1$\\
  [0.2em]
$\qquad  {\rm fermion}$  & $-$  & $\two$   & $- $ & $-$ &$-$  & $-$  & $+\frac52$ &$-\frac{1}{2}$ &   $-$\\ [0.2em]
 $S$    & $-$  & $\one$   & $0$ & $0$ & $0$ &$0$& $0$ &$+2$ &$1$ 
  \\ [0.2em]
$\qquad  {\rm fermion}$  & $-$   & $\one$  & $0$ & $0$  & $0$  & $0$& $0$ & $+1$ &$-$ \\ [0.2em]
 $ D^i\sim(\one,\two)$  & $-$   & $\two$  & $-$ & $-$   & $-$ &$+1$&  $-\frac52$ &$+\frac{3}{2}$ &$1$\\
  [0.2em]
$\qquad  {\rm fermion}$  & $-$   & $\two$ & $-$  & $-$  & $-$&$+1$  & $-\frac52$& $+\frac{1}{2}$ &$-$
\\[0.2em]
    \colrule
 \end{tabular}
 \end{center}
 \caption{Global symmetries with $N_c=5$.   }
 \label{tab:Composites}
 \end{table}
    
In the confinement process, the anomaly matching of global symmetries was suggested \cite{Hooft79}. Since we have the global symmetry \SU21, it is a question how an anomalous U(1) survives the confinement process. Not to worry about the perturbative effects of fermion representations, let us  gauge  the SU(2) part. For the global U(1), consider the chiral transformations on fermions by an angle $\theta$ of U(1).  If there is anomalies above the confinement scale, then due to the non-perturbative effects through instantons there is an  effective $\theta$ term for the non-Abelian SU(2). Use this chiral transformation of U(1) such that quark phases become 0.   But the anomalous $\theta$ term, generated non-perturbatively, survives the confinement process as discussed in axion physics. We will consider a global symmetry in the end, which will corresponds to an infinite (global) size of instanton and hence must be satisfied if a global SU(2) replaces the gauge SU(2).  In general, this argument leads to, ``if we consider a global non-Abelian group G then the anomalies of the forms G--G--G and U(1)--G--G should match above and below the confinement scale \cite{Hooft79}.'' For G=SU(2), there is no anomaly of the type G--G--G. So, in our case of G=SU(2), we consider matching the anomalies of U(1)$_{\rm global}-$SU(2)$_{\rm global}-$SU(2)$_{\rm global}$ above and below the confinement scale. But we need not consider U(1)$_{\rm global}^3$ anomaly since we cannot consider such a term from the above non-perturbative argument. 
 
To discuss the SU(5)  confinement, first let us consider U(1)$_{\rm global}$--SU(5)$^2_{\rm gauge}$ anomaly. 
As we commented above, there remains only one global U(1) for which we take $U(1)_{\bar{\Psi} }$.
 If we considered only matter fields, it is an anomaly free U(1)$_{AF}$,
\dis{
U(1)_{AF}=-U(1)_{\bar{\Psi} }.
}
But, we must consider  the R-symmetry U(1)$_R$ also.   These two U(1)'s are listed in Table \ref{tab:Composites}.  The U(1)$_R$ is   anomalous. The important point of R symmetry is not in the calculation of anomaly but in the calculation of superpotential in the SUSY case. The anomaly is a short distance effect, leading to (U(1)-charge)$F\tilde{F}$ vertex. As in the case of axion coupling to photon, we do not calculate (axion)-$F\tilde{F}$ even for the value at zero temperature, not by the composites such as proton and neutron but by the quarks. 
 For composites $\phi, \Phi_i$ and $S$, 
U(1)$_{AF}$ and  U(1)$_R$ are also listed in Table \ref{tab:Composites}. U(1)$_{AF}$, being SU(5)$_{\rm gauge}-$anomaly free, there is no constraint for the sum of  U(1)$_{AF}$ of composites to satisfy because there is no gauge symmetry we can consider below the confinement scale.

For the anomalies, we do not check U(1)$^3$ but   U(1)$_{\rm global}\times$SU(2)$_{\rm global}\times$SU(2)$_{\rm global}$ above and below the confinement scale. We must consider   U(1)$_{\rm global}\times$SU(2)$_{\rm gauge}\times$SU(2)$_{\rm gauge}$ first because `anomaly' is the anomaly in the renormalization of the gauge couplings. Then,  apply the result to the infinite spacetime to draw a conclusion on the global case.

 Since   U(1)$_R$ is anomalous, we consider it, 
checking only with the most fundamental particles. 
  Above the confinement scale, we consider doublets of global SU(2) only in $(\five,\two)$ from which we obtain U(1)$_R-$SU(2)$_{\rm global}-$SU(2)$_{\rm global}$ anomaly of $ (-\frac12)\times 5=-\frac52$ units.  As commented above,  below the confinement scale also, we must consider the most fundamenta  $(\five,\two)$ and obtain again $-\frac52$.
  
\begin{figure}[!t]
 \includegraphics[width=0.3\textwidth]{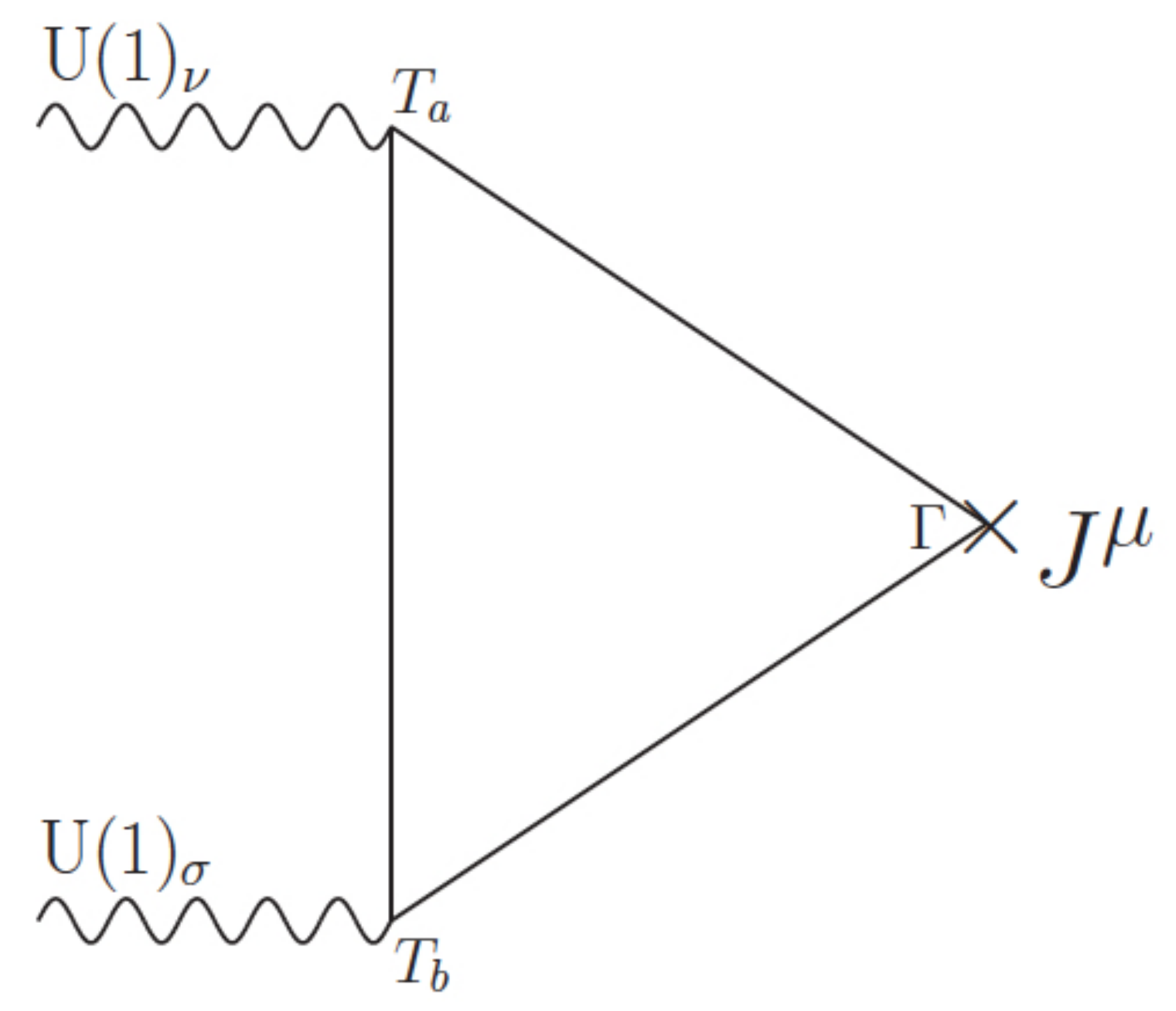}
\caption{The  $J_\mu$-U(1)$_{\rm gauge}$-U(1)$_{\rm gauge}$ anomaly.}\label{fig:U(1)3Anomalyi}
\end{figure}

Even if we keep two vertices attached by U(1) gauge bosons, the (global U(1))--(gauge U(1))--(gauge U(1)) triangle will give the anomaly of (global U(1) current)--$F_{\mu\nu}\tilde{F}^{\mu\nu}$ times group-theory factor $d_{\Gamma ab}=\Gamma\delta_{ab}$ as depicted in Fig. \ref{fig:U(1)3Anomalyi}. The spacetime part is a total derivative,
\dis{
\propto & \theta\frac{\epsilon^{\mu\nu\rho\sigma}}{4}(\partial_\mu A_\nu-\partial_\nu A_\mu)(\partial_\rho A_\sigma-\partial_\sigma A_\rho)=\theta\epsilon^{\mu\nu\rho\sigma}[(\partial_\mu A_\nu)(\partial_\rho A_\sigma)]\\
&=\partial_\rho[\theta \epsilon^{\mu\nu\rho\sigma}(\partial_\mu A_\nu)(  A_\sigma)]- [\theta \epsilon^{\mu\nu\rho\sigma}(\partial_\rho\partial_\mu A_\nu)]A_{\sigma} =\partial_\rho[\theta \epsilon^{\mu\nu\rho\sigma}(\partial_\mu A_\nu)(  A_\sigma)],
}
 and hence we can neglect the [global U(1)]--[gauge U(1)]$^2$ anomaly, depicted in Fig. \ref{fig:U(1)3Anomalyi}.  In fact, from the U(1) charges of Table \ref{tab:Composites}, one can easily show that the [global U(1)]--[gauge U(1)]$^2$ anomalies of above and below the confinement scale match, by redefining $U(1)_R+\frac{U(1)_{AF}+\ell(\two)}{2}$ as a new $U(1)_R$.

For the SU(2) discrete anomaly, the gauge boson attachement is not necessary and we have six(=even number) copies (together with $D_1$) of doublets above the confinement scale and two(=even number) copies (together with $D_1$) of doublets below the confinement scale. So, the Witten anomaly \cite{Witten82} is also matched. 

Therefore, we can consider the following superpotential below the confinement scale, consistently with our global symmetries,  \dis{
W=    M^2\phi  +\frac{N_c(N_c^2-1)}{32\pi^2} \mu_0^2S\left(1-a\log\,\frac{\Lambda^3}{S \mu_0^2}\right)+  bM\Phi_i D^i ,
  \label{eq:Wbelow}
}
where   $\Lambda$   is the definition for the gaugino condensation scale leading to the Veneziano-Yankielowicz composite $S$  \cite{Veneziano82}.  [Note that  ours as an effective one still respects the R symmetry in contrast to the phenomenological Polonyi potential $W=m^2 z+{\rm constant}$ \cite{PolonyiJ77}.]  In the superpotential, we   considered only $ M^2\phi\equiv M_1^2\phi _1 +M_2^2\phi_2$ since the other linear combination does not appear at all.   
The parameter $M^2$ can be calculated from original superpotential
 \dis{
 \frac{\lambda_0}{5!}\,\bar{\Psi}^{\alpha\beta}\bar{\Psi}^{\gamma\delta}\bar{\psi}_1^{\epsilon}\,\epsilon_{\alpha\beta\gamma\delta\epsilon}\to \lambda_0 \mu_0^2\,\phi,\nonumber
 }
 leading to $M^2\simeq \lambda_0\Lambda^2$. So, the SUSY conditions for the composites are
\dis{
&\frac{\partial W}{\partial \phi}=0: ~M^2 =0,\\ 
& \frac{\partial W}{\partial \Phi_i}=0: ~   D^i =0, \\ 
& \frac{\partial W}{\partial D^i}=0: ~\Phi_i =0,\\ 
&\frac{\partial W}{\partial S}=0: ~ \mu_0^2\left( 1+a-a\,\log\,\frac{\Lambda^3}{S\mu_0^2}\right)=0, \label{eq:ORai}
}
where the first equation cannot be satisfied. 
 The SUSY conditions determine $\langle\Phi_i\rangle=0$ and  $\langle D^i\rangle=0$.
One can add one more pair of doublets $D_2$ and $D_3$ to break the global symmetry \SU21\, to a global U(1). At this stage, it is difficult to break the remaining U(1) by another term in the superpotential. Let the remaining global symmetry be manifested by the phase of $\phi$.  
The VEV $\langle S\rangle$ is determined from the SUSY condition at $ (\Lambda^3/\mu_0^2)e^{-(1+a^{-1})} $. Equations in (\ref{eq:ORai}) show that the SUSY breaking is not by  $\langle S\rangle$ but through the $M^2$ term due to the O'Raifeartaigh mechanism \cite{ORaifeartaigh}.
 
One of the popular scenarios for SUSY breaking is the gaugino condensation triggering the $F$-term(s) of singlet chiral field(s) $z$. The gauge kinetic function $f(z)$ in $\int d^2\vartheta\, f(z)W^\alpha W_\alpha$  can take a form $f(z)=1+ (z/M)$ in this case  \cite{GauginoFGN83} and $z$ must carry $R$ charge 0. From matter fields of $\EE8$, we encountered that there does not appear a singlet field through compactificatioin with all U(1) charges being zero \cite{Munoz93,Huh09}. Therefore, a candidate can be only from the antisymmetric tensor field $B_{MN}$, which was suggested in \cite{DIN85,DRSW85}. It is commented that the only possible  bilinear of Majorana-Weyl gluinos in 10 dimensions is Tr\,$\bar{\chi}\Gamma_{MNP}\chi\,(M,N,P=1,\cdots,10)$ and hence the coupling can be $B^{MNP}$Tr\,$\bar{\chi}\Gamma_{MNP}\chi$ \cite{DRSW85}. One needs a large value, $\gtrsim 10^{13\,}\gev$, for gluino condensation becasuse of  gluino of dimension  $\frac32$.  With our  $\phi$ of  dimension 1, the same amount of SM SUSY breaking is obtained by our $M$ at $5\times 10^{10\,}\gev$.  Therefore, if  SUSY breaking between $10^{13\,}\gev$ and $5\times 10^{10\,}\gev$ is needed then our mechanism is useful in raising the superpartner masses above TeV.
 
Since $V$ is bounded by the SUGRA correction, 
$\phi$ get a VEV of order $M_P.$ 
So the SUGRA corrections are not negligible. 
So the Polonyi problem (=moduli problem) is there as is well known.

 \section{Effects of supergravity}
 
 The SM particles acquire the SUSY breaking effects  from the hidden sector by   supergravity effects, proportional to the gravitino mass \cite{vanPoyen83}. The scalar partners of quarks and leptons, and the fermionic partner of gauge bosons  acquire masses at the order the gravitino mass $m_{3/2}^2\simeq |F|^2/3\Mp^2$ where $F$ is the SUSY breaking $F$-term \cite{Zumino77}, and for the minimal Kahler potential $K= |\phi |^2+\cdots$
 \dis{
F = DW = (\partial_\phi W) + (\partial_\phi K) W / M_P^2
= M^2 + \phi^* M^2 ( \phi +N_c \Lambda^3 e^{i\alpha} / M^2) / M_P^2 . 
}
  In this case, the vacuum energy $V$ is  \cite{vanPoyen83},
 \dis{
\frac{ V}{\exp(|\phi|^2/\Mp^2)}= |F|^2-\frac{3|W|^2}{\Mp^2}&\\
= M^4\left(1+\frac{2|\phi|^2}{\Mp^2}+ \frac{|\phi|^4}{\Mp^4}\right)&+\frac{\Lambda^6 N_c^2|\phi|^2}{\Mp^4} + \frac{M^2\Lambda^3 N_c(\phi e^{-i\alpha} +\phi^*e^{i\alpha} )}{\Mp^2} - \frac{3}{\Mp^2} \left|M^2\phi+N_c\Lambda^3 e^{i\alpha} \right|^2 ,\label{eq:V}
 }
 where  we considered only vacuum values of $M^2$ and $S$ terms in $W$ and $\alpha=0$ or $\pi$.  
Here, we note that the U(1)$_R$ charges of $\phi$ (both $\phi_1$ and $\phi_2$) and $S$ are identically 2, viz. Table \ref{tab:Composites}, implying that both VEVs  $\langle \phi\rangle$ and $\langle S\rangle$ break U(1)$_R$ down to $\Z_2$. Therefore, the phases of $\langle \phi\rangle$ and $\langle S\rangle$,  $\delta$ and $\alpha$, have the same periodicity: $\delta-\alpha=$constant. Therefore, the last two term in Eq. (\ref{eq:V}) does not have a $\delta$ dependence. $\delta$ is a flat direction.
\color{black}
The dynamically generated superpotential of gluino condensation corresponds to the $N_f=0$ case in SQCD, \ie 
the Affleck-Dine-Seiberg superpotential ${\cal W}=N_c \Lambda^3$ \cite{Affleck84} where $ \Lambda^3$ carries two units of $R$ charge. 
 
\begin{figure}[!h]
 \includegraphics[width=0.6\textwidth]{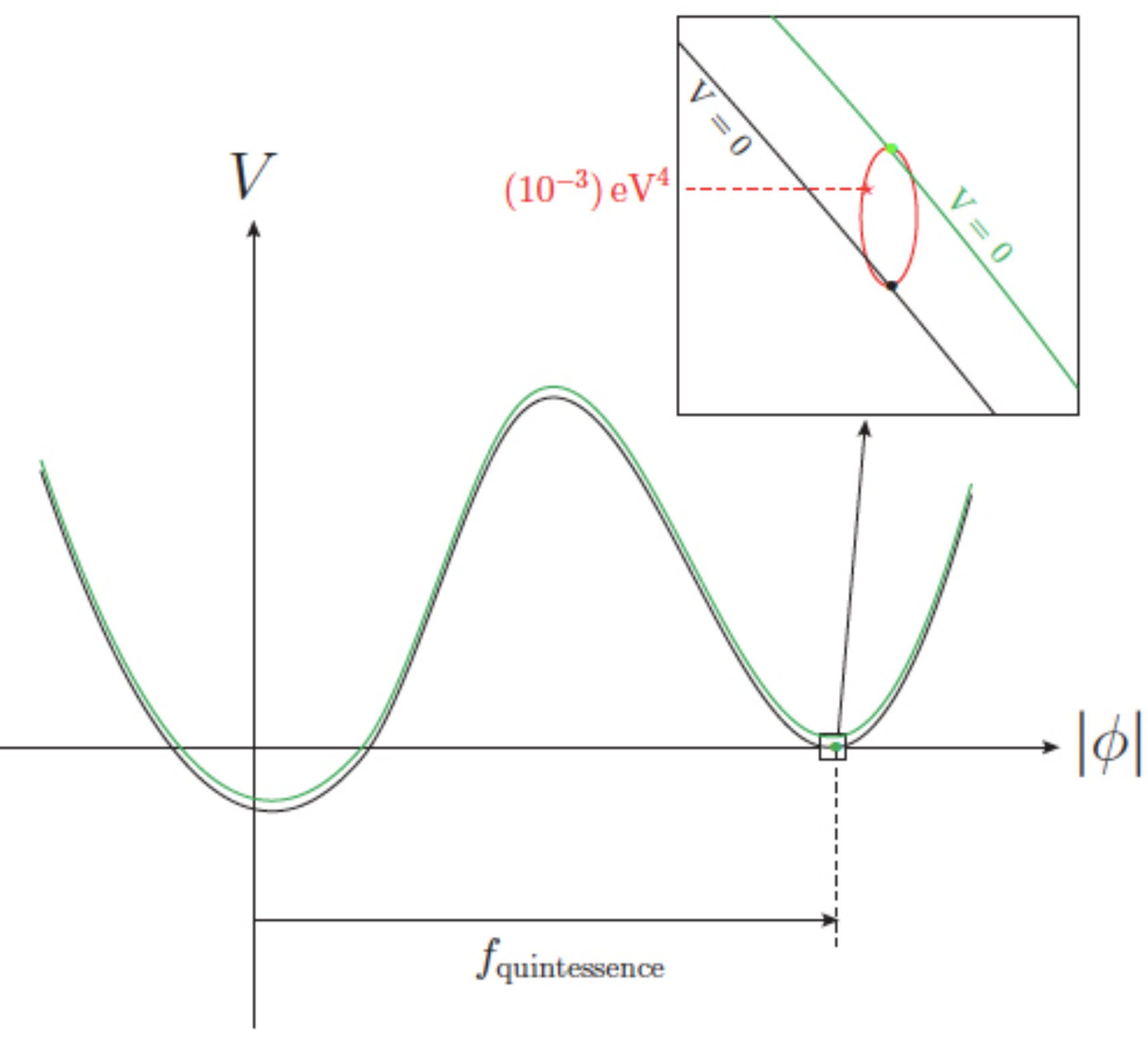}
\caption{The cosmological constant as a function of $\phi$, one of which having $V=0$ is in the square. Solutions of Eq.  (\ref{eq:vSolution}) are the points with $dV/dv=0$.  Each sign of  Eq.  (\ref{eq:V}) for $\alpha=0$ and $\pi$ are shown as the black and green curves, by bringing  the parity reflected curve ($\alpha=\pi$) near the other curve ($\alpha=0$).   In the inset, the square is amplified with the black and the green bullets corresponding to  $\alpha=0$ and $\pi$, respectively. The red star corresponds to the current quintessential point $\phi=|\phi|e^{i\delta}$  \cite{KimPLB03}.}\label{fig:Vphi}
\end{figure}

Let us determine $\langle\phi\rangle\equiv v e^{i\delta}$ such that $V=0$ for two vacua of gluino condensates. To illustrate solutions, let us keep the leading terms in the $(1/\Mp^2)$ expansion to obtain the minimization condition
 \dis{
 M^4\, v^3 -4M^2\Lambda^3 N_c     \cos(\delta-\alpha)\, v^2 -3 N_c^2 \Lambda^6 v-2\Mp^2 M^2\Lambda^3 N_c\cos(\delta-\alpha)=0 .\label{eq:vSolution}
 } 
 Equation (\ref{eq:vSolution}) must be satisfied together with $V=0$  
which are achieved by fine-tuning of parameters. Namely, setting the cosmological constant at the minimum as zero is by fine-tuning.   In Fig. \ref{fig:Vphi}, one such minimum is chosen as a square.  With this fine-tuning, we set the vacuum energy of these two  vacua at zero, which are the black and green bullets in the inset of Fig. \ref{fig:Vphi}. Between  the black and green bullets, the vacuum energy is positive.
 
In Fig. \ref{fig:Vphi},   we sketch a rough idea of the vacuum energy dependence on the composite $\phi$. 
 The curvature at the $v$ direction is known as the Polonyi problem \cite{Kolb83}. In our case the maximum point in the center is about $\Lambda^4$ and $f_{\rm quintessence}$ is about $\lesssim \Mp$, and a rough estimate of the mass of real $\phi$ is  $\gtrsim (10^{12\,}\gev)^2/\Mp\approx 10^{4\,}\gev$. So, the Polonyi problem can be evaded for a little bit larger value of $\Lambda$ compared to $\mu_0$.

 \section{Quintessential axion as dark energy}
 
 In addition, we need a mechanism to break the global U(1) explicitly because gravity does not allow an exact global symmetry \cite{RevGlobal17}. If the magnitude of breaking is tiny,$\sim (10^{-3}\eV)^4$, and the radius of the circle in the inset of Fig. \ref{fig:Vphi} is large, $\lesssim \Mp$, it is the quintessential axion \cite{Carroll98,KimPLB03}. In  Fig. \ref{fig:Vphi}, we depict the anticipated vacuum energy as a function of $\phi$. The $\Z_2$ vacua from the gaugino condensation are denoted as the black and green bullets, which are determined by fine-tuning as commented below Eq. (\ref{eq:vSolution}).  In the inset, the quintessential field is along the red circle with radius $f_{\rm quintessence}$ and the star descibes the current magnitude of dark energy.

 With the field content present in Table \ref{tab:Composites}, it is possible to realize a quitessential axion.
 The complex $\phi$ accompanies a global U(1) symmetry, as commented below Eq. (\ref{eq:ORai}).   The height at the star is small because of the composite nature of our $\phi$ and the difficulty of breaking SU(2) global symmetry.   In other words, it is commented below Eq. (\ref{eq:V}) that the phase is a flat direction at this level of requiring U(1)$_R$ charge 2 for the superpotential.   
 The composite fields $\phi_1, \phi_2$ and $\Phi_i$ of Eq.  (\ref{eq:Comp}) and $D_i$ are the fields we consider at low energy. In terms of these, we cannot consider a superpotential term. But a D-term can be considered from $\phi^* \Phi_i D_j\epsilon^{ij}$ which can be given  in terms of SU(5) fields as\footnote{For $\phi$, we illustrate in terms of $\phi_2$. $\phi_1$ would have given  the same power of mass suppression.}
 \dis{
 \frac{1}{\tilde{M}^4}\int d^2\vartheta d^2\bar{\vartheta}\,\left( \frac14\bar{\Psi}^{\alpha\beta}\psi_{2\alpha}^i \psi_{2\beta}^j\epsilon_{ij}\right)(\bar{\psi}_1^\alpha {\psi_{2\,\alpha\,k}} )^*(D_l)^*\epsilon^{kl}+{\rm h.c.}
 }
where $\tilde{M}$ is some Planck  scale mass parameter. 
From Eq. (\ref{eq:Wterms}), the F-term of $(D_i)^*$ is $\bar{\psi}_1^\alpha {\psi_{2\,\alpha\,k}} $, and integrating out $d^2\bar\vartheta$ gives
 \dis{
 \frac{|\bar{\psi}_1^\alpha {\psi_{2\,\alpha\,k}}|^2}{\tilde{M}^4}\int d^2\vartheta  \,\left( \frac14\bar{\Psi}^{\alpha\beta}\psi_{2\alpha}^i \psi_{2\beta}^j\epsilon_{ij}\right) +{\rm h.c.}\label{eq:effW}
 }
But Eq. (\ref{eq:effW}) is of the form of Eq. (\ref{eq:Wbelow}), and hence it cannot generate the height of a quintessental axion potential.
In terms of an effective superpotential, we cannot generate  the potential for the quintessental axion.
Since our global symmetry has originated from the anomaly, one can consider the QCD anomaly first. But it is too big for the cosmological constant. Anyway, it must be reserved for the invisible axion \cite{KimPLB84,KimKyaeNam17}. The next obvious contribution is the SU(2)$_W$ anomaly which is sufficiently small,
 \dis{
 \sim e^{-2\pi/\alpha_2}\vew^4\sim 10^{-81.6}\vew^4\sim 10^{-5}(10^{-3}\eV)^4.
 }
But, it is a bit small for  dark energy of a quinessential axion. Note that with approximate global symmetries allowed by discrete symmetries in string compactification, there can be a contribution as discussed in Ref. \cite{Kimjhep00}. 
 
 But, when we consider gravity, we should not have a global symmetry. String compactification must allow U(1)$_R$ breaking terms.  Indeed,  this interesting spectrum (\ref{eq:qrepresentations}) in the hidden sector   was obtained before from string compactification \cite{Huh09}, where   the visible sector flipped SU(5) model with three families are realized \cite{Barr82,DKN84}. Recently, a successful $\Z_{4R}$ symmetries needed for proton longevity \cite{Lee11,KimZnR} has been assigned successfully in this model \cite{KimPRD19}. To illustrate the example, in Table II   we list the SGUT non-trivial spectra from $\Z_{12-I}$ orbifold compactification \cite{Huh09}, where many vector-like pairs of $(\five\oplus \fiveb)$ of \cite{Huh09} are not listed. U(1)$_R$ charges as in Ref.  \cite{KimPRD19} are listed in the 4th column.  Removing the vector-like representation $F_3'\oplus F_4'$ just below the string scale by a superpotential term $\sim F_3' F_4'$, we obtain the spectra presented in Eq. (\ref{eq:qrepresentations}). The model contains many SM singlet fields $\sigma_{I}$ whose  U(1)$_R$ charges can be found  in Ref.  \cite{KimPRD19}. Some singlets carry negative U(1)$_R$ charges and we can consider superpotential terms of the form
 \dis{
 W'\propto \frac{1}{\tilde{M}^{n+3}} \,\left( \frac14\bar{\Psi}^{\alpha\beta}\psi_{2\alpha}^i \psi_{2\beta}^j\epsilon_{ij}\right)(\bar{\psi}_1^\alpha {\psi_{2\,\alpha\,k}} ) (D_l)  \epsilon^{kl}\left(\sum_{I=1}^n\sigma_{a_1} \cdots \sigma_{a_n} \right)
 }
 such that 
 \dis{
  \sum_{I=1}^n Q(\sigma_{a_I})=-2 .\label{eq:Qcharge}
 }
  The GUT scale VEVs $\sigma_{a_I}$ break the U(1)$_R$ symmetry  and indeed there is no global symmetry when we consider gravity. For the quintessential axions to be realized from string compactification, the model is constructed such that it is not possible to allow VEVs of scalars satisfying Eq. (\ref{eq:Qcharge}).
  
\begin{table}[t!]
\begin{center}
\begin{tabular}{@{}lc|cc|c|cccccc|ccc@{}} \toprule
 &State($P+kV_0$)&$~\Theta_i~$ &${\bf R}_X$(Sect.)&$Q_R$ &$Q_1$&$Q_2$ &$Q_3$ &$Q_4$ &$Q_5$ &$Q_6$ &   $Q_{18}$& $Q_{20}$& $Q_{22}$
  \\[0.1em] \colrule
 $T_1'$  & $(0^5;\frac{-1}{6}\,\frac{-1}{6}\,\frac{-1}{6})(
 \underline{-1\,0^3}\,0;\frac{+1}{4}\,\frac{-1}{4}\,\frac{+1}{2})'$&$0$ &$\tenb_{0}'(T_1^0)_R$&$+1$ &$-2$ & $-2$ & $-2$ & $0$ & $+18$ & $+6$ &   $-1$ & $+1$& $-1$ \\
   & $(0^5;\frac{-1}{6}\,\frac{-1}{6}\,\frac{-1}{6})(
 \underline{\frac{+1}{2}\,\frac{+1}{2}\,\frac{-1}{2}\frac{-1}{2}}\,0;\frac{+1}{4}\,\frac{-1}{4}\,\frac{+1}{2})'$&  & &  & &   &   &  & & &    & &  \\[0.2em]
 $F_1'$  & $(0^5;\frac{-1}{6}\,\frac{-1}{6}\,\frac{-1}{6})(
 \underline{+1\,0^3}\,0;\frac{+1}{4}\,\frac{-1}{4}\,\frac{+1}{2})'$&$0$ &$(\five',\two')_0(T_1^0)_R$&$+1$ &$-2$ & $-2$ & $-2$ & $-12$ & $+18$ & $+6$ &   $-1$ & $+1$& $-1$ \\
   & $(0^5;\frac{-1}{6}\,\frac{-1}{6}\,\frac{-1}{6})(
 {0\,0\,0\,0}\,-1;\frac{+1}{4}\,\frac{-1}{4}\,\frac{-1}{2})'$&  & &  & &   &    & & & &  & &  \\ 
   & $(0^5;\frac{-1}{6}\,\frac{-1}{6}\,\frac{-1}{6})(
 \underline{\frac{+1}{2}\,\frac{-1}{2}\,\frac{-1}{2}\,\frac{-1}{2}}\,\frac{-1}{2};\frac{-1}{4}\,\frac{-3}{4}\,0)'$&  & &  & &   &  & & & &  & &  \\ 
   & $(0^5;\frac{-1}{6}\,\frac{-1}{6}\,\frac{-1}{6})( {\frac{+1}{2}\,\frac{+1}{2}\,\frac{+1}{2}\,
     \frac{+1}{2}}\,\frac{-1}{2};\frac{+3}{4}\,\frac{+1}{4}\,0)'$&  & &  & &   &  & & & &  & &  \\[0.2em]
 $F_2'$  & $(0^5;\frac{-1}{6}\,\frac{-1}{6}\,\frac{-1}{6})( \underline{\frac{-1}{2}\,\frac{+1}{2}\,  
    \frac{+1}{2}\,\frac{+1}{2}}\,0;\frac{+1}{4}\,\frac{-1}{4}\,\frac{-1}{2})' $&$0$ &$ \fiveb_{0}'(
      T_1^+)_R$&$0$ &$+4$ & $-4$ & $0$ & $+12$ & $-18$ & $-6$  & $+1$ & $+1$& $+1$ \\
   & $(0^5;\frac{-1}{6}\,\frac{-1}{6}\,\frac{-1}{6})(0\,0\,  
    0\,0\,\frac{+1}{2};\frac{-1}{4}\,\frac{+1}{4}\,0)' $
   &  & &  & &  &  & & & &  & &  \\ [0.2em]
 $F_3'$  & $((\frac{+1}{6})^5;\frac{-1}{3}\,\frac{+1}{3}\,0)(
   \underline{\frac{-5}{6}\,\frac{+1}{6}\,\frac{+1}{6}\,\frac{+1}{6}}\,\frac{+1}{2};\frac{-1}{12}\, 
    \frac{-1}{4}\,0)' $&$0$ &$ \fiveb_{ -5/3}'(T_1^+)_R$&$-4$ &$-4$ & $+4$ & $0$ & $-4$ & $-6$
      & $-6$  & $-1$ & $-1$&$+1$ \\
   & $((\frac{+1}{6})^5;\frac{-1}{3}\,\frac{-1}{3}\,0)(
  {\frac{+1}{3}\,\frac{+1}{3}\,\frac{+1}{3}\,\frac{+1}{3}}\,0;\frac{-1}{12}\,\frac{+1}{4}\,\frac{-1}{2})'$
   &  & &  & &  &  & & & &  & &  \\ [0.2em]
$F_4'$  & $( (\frac{-1}{6})^5;\frac{-1}{3}\,0\,\frac{+1}{3})(
  (\underline{\frac{+5}{6}\,\frac{-1}{6}\,\frac{-1}{6}\,\frac{-1}{6}}\,\frac{-1}{2}; \frac{+1}{12}\,\frac{+1}{4}\,0)'$&$0$ &$\five_{+5/3}'(T_7^-)_R$&$0$ &$-4$ & $0$ & $+4$ & $+4$ & $-6$ & $+6$   & $-1$ & $-1$& $+1$ \\
   & $((\frac{-1}{6})^5;\frac{-1}{ 3}\,0\,\frac{+1}{3})(
 {\frac{+1}{3}\,\frac{-2}{3}\,\frac{-2}{3}\,\frac{-2}{3} }\,-1;\frac{-5}{12}\,\frac{-1}{4}\,\frac{-1}{2})'
$&  & &  & &  &  & & & &  & &  \\ [0.2em]
  \botrule
 \end{tabular} \label{tab:SMqn}
\end{center}
\caption{U(1) charges of SU(5)$'$ fields. Here, U(1)$_R$ is defined by $Q_R=\frac{1}{2}(Q_1+Q_2+Q_3)+\frac13 Q_6+2Q_{20}$.
}
\end{table}

\section{Conclusion}
We showed that a chiral representation,  $\tenb\, \oplus\, \fiveb\, \oplus\,  2\cdot \five$   in supersymmetric SU(5), breaks  global supersymmetry dynamically. This mechanism  of supersymmetry  breaking can generate smaller mass scales in physics if appropriate discrete symmetries are provided.  We obtained this needed spectra from the hidden sector of heterotic string as shown in a  $\Z_{12-I}$ orbifold compactification. There can be many useful applications of this dynamical breaking of supersymmetry. Firstly, the little hierarchy of factor $r$ in particle physics, \ie supersymmetry  appearing above $r\, \tev$, is obtained if SU(5) confines at $\gtrsim\sqrt{r}\cdot 5\cdot 10^{10\,}\gev$. 
Second, dark energy in the Universe can be attributed to the composite nature of $\phi$ arising from the SU(5) confinement at field theory framework but realization from string compactification needs a judicious symmetry breaking pattern.
  
\acknowledgments{J.E.K. is supported in part by the National Research Foundation (NRF) grant   2018R1A2A3074631, and B.K. is supported in part by the NRF grant   2016R1D1A1B03931151.}  


\end{document}